\documentclass[pdftex,twocolumn,epjc3]{svjour3}          % twocolumn

\RequirePackage[T1]{fontenc}

\smartqed  % flush right qed marks, e.g. at end of proof

\RequirePackage{graphicx}
\RequirePackage{mathptmx}      % use Times fonts if available on your TeX system
\RequirePackage{flushend}
\RequirePackage[numbers,sort&compress]{natbib}
\RequirePackage[colorlinks,citecolor=blue,urlcolor=blue,linkcolor=blue]{hyperref}

\usepackage{amssymb, graphics, setspace}
\usepackage{pgfplots}
\usepackage{textcomp}
\usepackage[caption=false]{subfig}
\usepackage{amsmath}
\usepackage{commath}
\usepackage{graphicx,bm}
\usepackage{url}
\usepackage{float}
\usepackage{textcomp}%\usepackage[labelfont=bf]{caption}
\usepackage{verbatim}
\journalname{Eur. Phys. J. A}

\begin{document}

\title{Shape of Proton and the Pion Cloud}

\subtitle{}

\author{L\'aszl\'o Jenkovszky\thanksref{e1,addr1}
        \and
        Istv\'an Szanyi\thanksref{e2,addr2}
        \and
        Chung-I Tan\thanksref{e3,addr3}
}

\thankstext{e1}{jenk@bitp.kiev.ua}
\thankstext{e2}{sz.istvan03@gmail.com}
\thankstext{e3}{chung-i$\_$tan@brown.edu}

\institute{Bogolyubov Institute for Theoretical Physics (BITP),\\
	Ukrainian National Academy of Sciences \\14-b, Metrologicheskaya str.,
	Kiev, 03680, UKRAINE\label{addr1}
          \and
          Uzhgorod National University, \\14, Universytets'ka str.,  
          Uzhgorod, 88000, UKRAINE\label{addr2}
          \and
          Department of Physics, Brown University, Providence, RI 02912, USA\label{addr3}
}


\maketitle

\begin{abstract}
Proton-proton differential and total cross sections provide information on the energy dependence of proton  shape and size.    We show that the deviation from exponential behavior of the diffraction cone observed near $t=-0.1$ GeV$^2$,  (so-called break),  both at the ISR and the LHC follows from the $t$-channel two-pion loop contributions, imposed by unitarity. By using a simple Regge-pole model, we extrapolate the "break" from the ISR energy region to that of the LHC. This allows us to answer two important questions: 1)~To what extent is the "break" observed recently at the LHC a "recurrence"  of that seen at the ISR (universality)? 2)~What is the relative weight of two-pion effect to the vertex coupling (Regge residue) compared to expanding size (pomeron propagator) in producing the "break"? We find that the effect comes both from the Regge residue (proton-pomeron coupling) and from the Regge propagator. A detail analyses of their balance, including the correlation between the relevant parameters is presented.  
\end{abstract}

\section{Introduction} \label{s1}

Following TOTEM's recent results \cite{TOTEM8} on the deviation from the exponential behavior of the diffraction cone at low-$|t|$ of the $pp$ differential cross section at $8$ TeV, and anticipating their new measurements at $13$ TeV, announced recently \cite{TOTEM13}, we find appropriate to update our earlier results and revise the physics behind the phenomenon. 

The diffraction cone of high-energy elastic hadron scattering deviates from a purely exponential dependence on $t$ due to two structures clearly visible in proton-proton scattering: the so-called "break" (in fact, a smooth curve) near $t=-0.1$ GeV$^2$, whose position is nearly independent of energy and the prominent "dip" -- diffraction minimum, moving slowly (logarithmically) with $s$ towards smaller values of $|t|$, where $s$ and $t$ are the Mandelstam variables. 

\begin{figure*}
	\centering
	\includegraphics[width=0.41\textwidth]{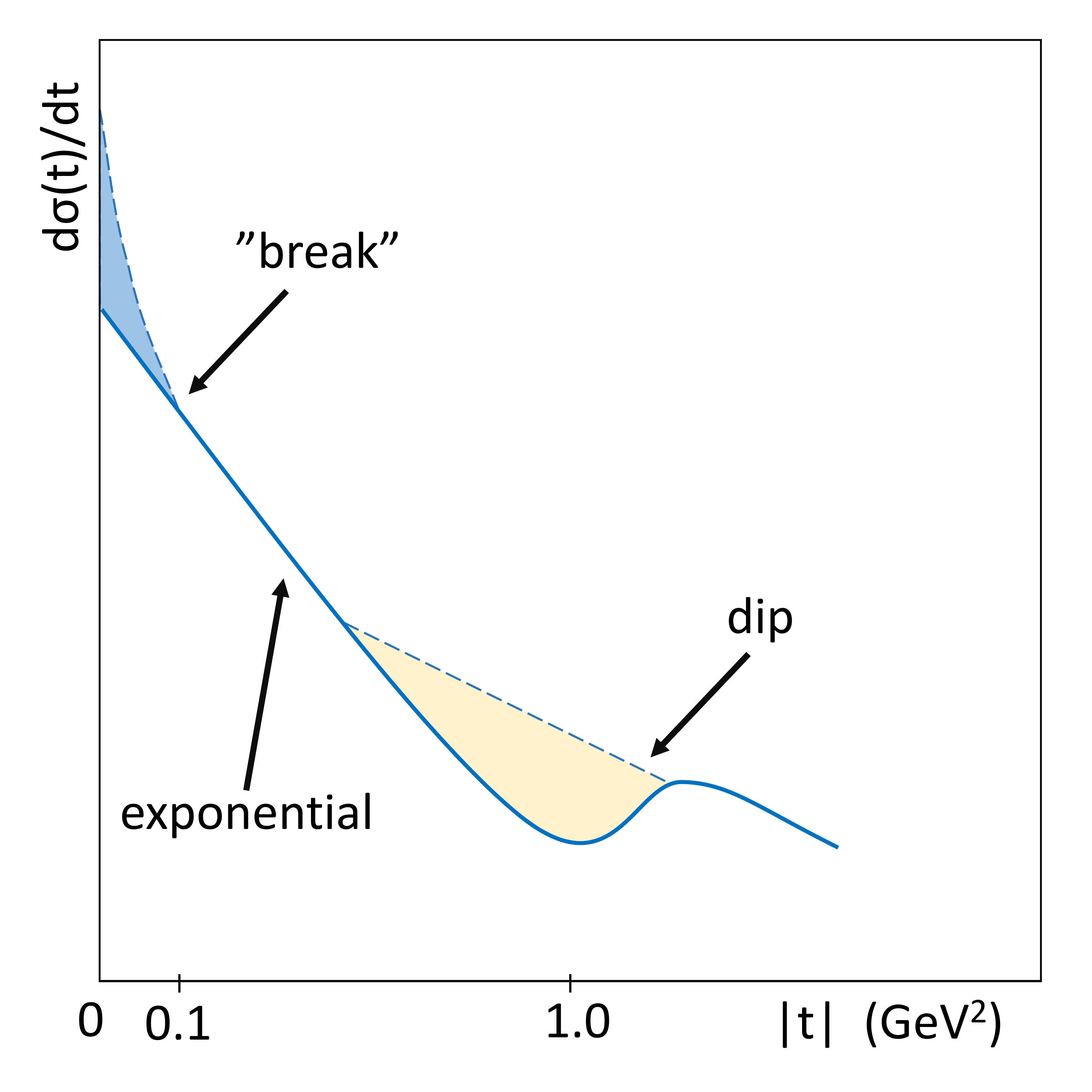}
	\includegraphics[width=0.41\textwidth]{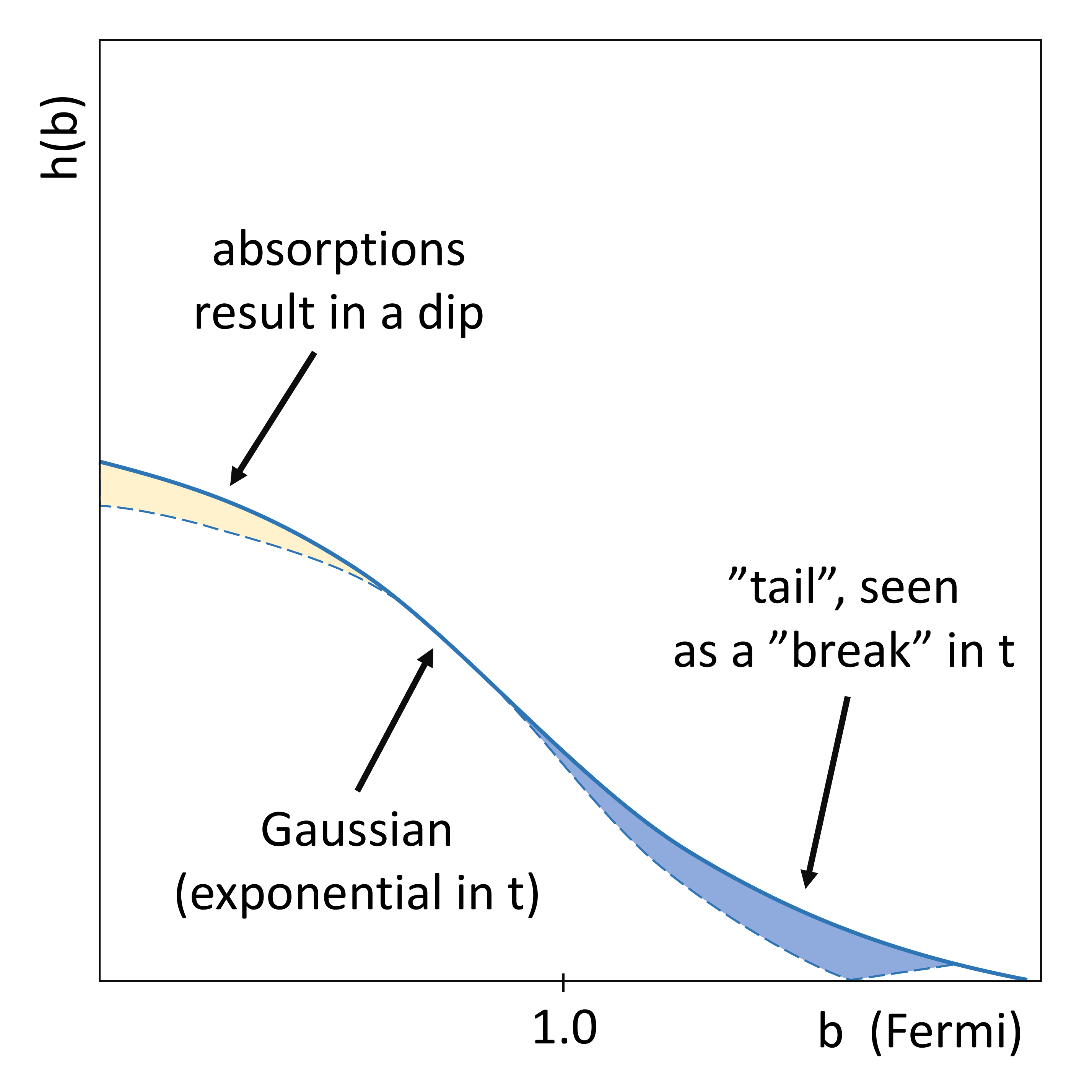}
	\caption{Schematic (qualitative) view of the "break", followed by the diffraction minimum ("dip"), shown both as function in $t$ and its Fourier transform (impact parameter representation), in $b$. While the "break" reflects the presence of the pion cloud around the nucleon at the outer edge of an expanding disk  in $b$, the dip results from absorption corrections, suppressing the impact parameter amplitude at small $b$.}
	\label{Fig:1}
\end{figure*}

\begin{figure*}
	\centering
	\includegraphics[width=1\textwidth]{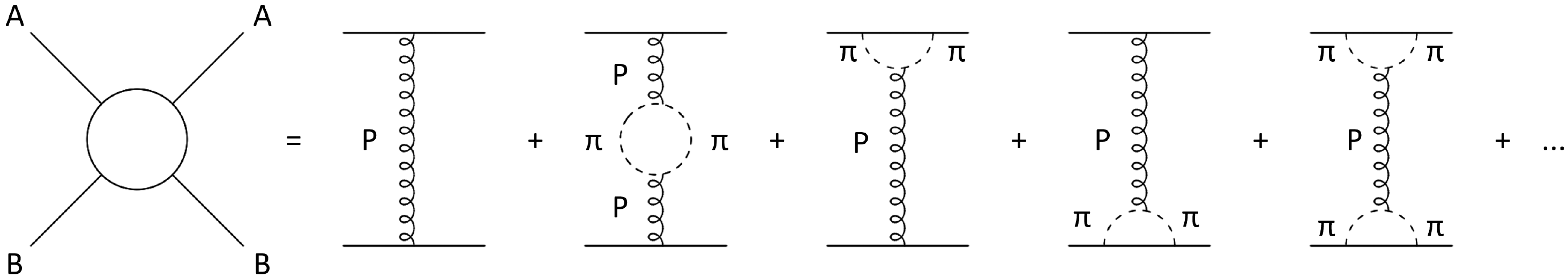}
	\caption{Diagram for elastic scattering with $t$-channel exchange containing a branch point at $t=4m_{\pi}^2$.} 
	\label{Fig:Diagram}
\end{figure*}

The physics of these two phenomena are quite different.
As illustrated in Figure~\ref{Fig:1}, the "break" is likely a reflection due to the "pion cloud", which controls  the ``static size" of nucleon. This effect, first observed in 1972 at the ISR, was interpreted \cite{LNC, AG, C-I1, C-I2} as the manifestation of $t$-channel unitarity, generating a two-pion loop in the cross channel, Figure~\ref{Fig:Diagram}, and was  referred to  by Bronzan \cite{Bronzan} as the ``fine structure" of the pomeron. This has been examined more closely  in a number of papers \cite{BPM, Brazil, RPM}. In Ref.~\cite{BPM} the "break" was fitted by a relevant form factor (residue function) in the Regge-pole scattering amplitude.
The dip (diffraction minimum), on the other hand, is generally accepted as a consequence of $s$-channel unitarity or absorption corrections to the scattering amplitude. As such, dip location reflects the ``size" of proton, moving towards smaller $|t|$ values as the total cross section increases with energy.

\begin{figure*}
	\centering
	\includegraphics[width=0.45\textwidth]{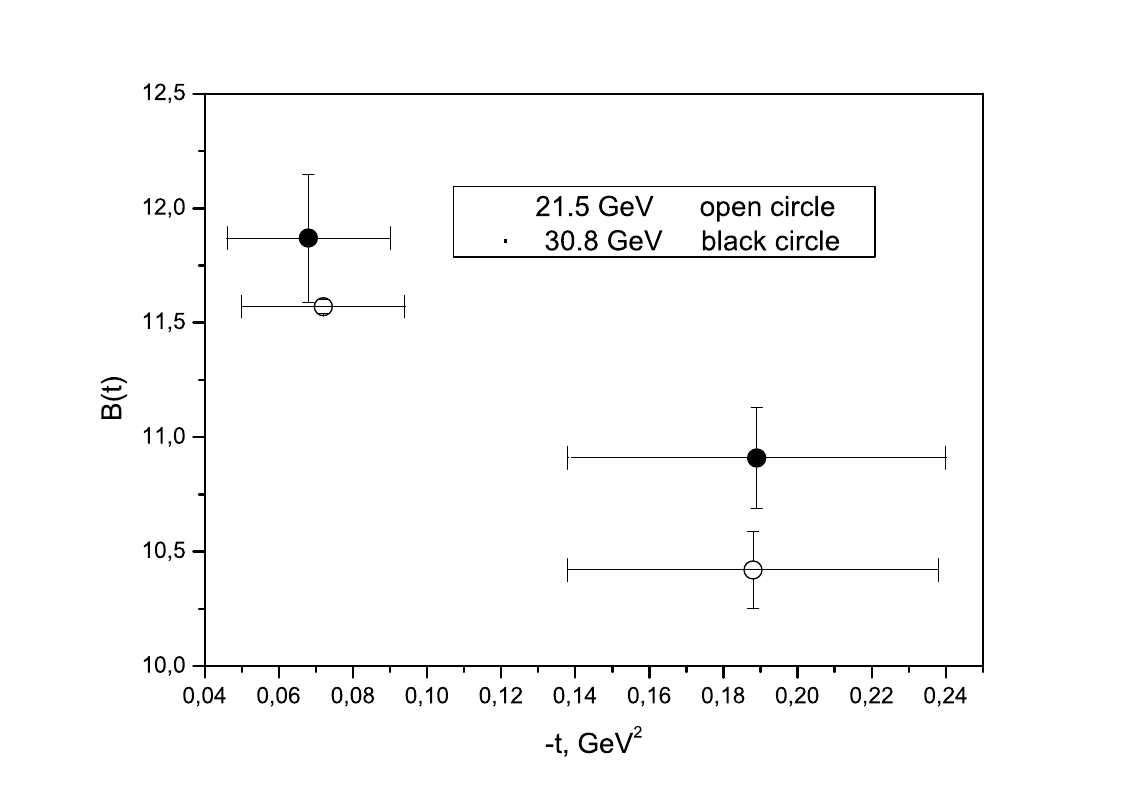}%
	\quad
	\includegraphics[width=0.45\textwidth]{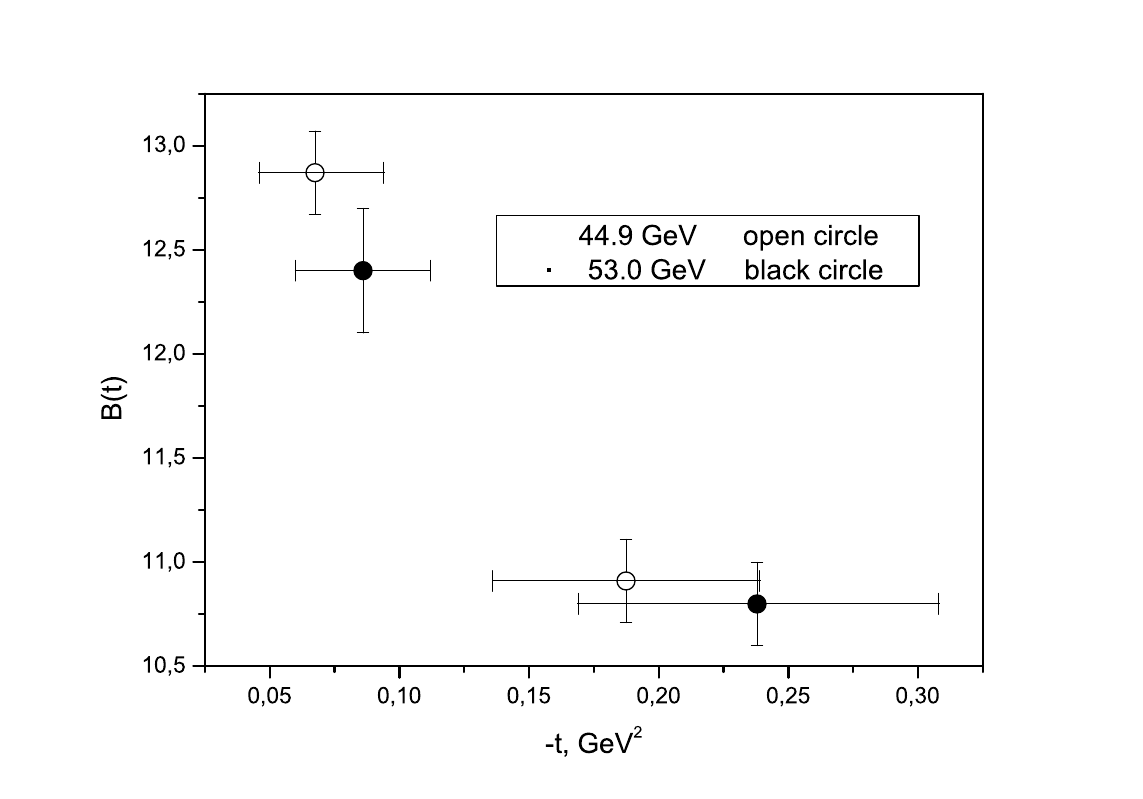}%
	\caption{Local slopes $B(t)$ calculated for the ISR data at 21 and 30 GeV (left) and 45 and 53 GeV (right) \cite{Bar}.}
	\label{Fig:slopes}
\end{figure*}

To quantify the departure from the linear exponential behavior, one can parametrize differential cross section as~\cite{TOTEM8, TOTEM13} 
\begin{equation}
|A_N|^2=ae^{B\, t}\rightarrow ae^{B(t)}=ae^{b_1 t+b_2 t^2+b_3 t^3+...},
\end{equation}
with coefficients $a$ and $b_i$ fitted to the data. This non-linear effect was confirmed by recent measurements by the TOTEM Collaboration at the CERN LHC, first at $8$ TeV (with a significance greater then 7$\sigma$) \cite{TOTEM8} and subsequently at $13$ TeV \cite{TOTEM13}. 
At the ISR the "break" was illustrated by plotting the local slope $B(t)$ for several $t$-bins at fixed values of $s$. As shown in Figure~\ref{Fig:slopes} from \cite{Bar} the local slope drops around $-0.1$ GeV$^2$ within the interval $\Delta t\approx0.1$ GeV$^2$  by about $\Delta B=2$ GeV$^2$. 

At the LHC the effect appears of the same order of magnitude and is located near the same value of $t$. Different from the ISR \cite{Bar}, TOTEM quantifies the deviation from the exponential by normalizing the measured cross section to a linear exponential form, (see Eq.~(\ref{Eq:norm}) below). For the sake of completeness we will exhibit this ``break effect"  both in the normalized form and for $B(t)$.    

The "break" (in fact a smooth deflection of the linear exponential) of the cone, has a relatively narrow location around $-t \approx 0.1 \pm 0.01$ GeV$^2,$ both at the ISR and the LHC energies, whereupon it recovers its exponential shape, followed by the dip, whose position is energy-dependent.

The new LHC data from TOTEM at $8$ TeV confirm the conclusions \cite{LNC, AG, C-I1, C-I2} about the nature of the break and call for a more detailed analysis and better understanding of this phenomenon. The new data triggered intense theoretical work in this direction \cite{JL, Brazil, RPM}, but many issues still remain open.
%{\bf The curvature of $B(t)$, both at the ISR and the LHC is concave.} 
While the departure from a linear exponential was studied in details both at the ISR and LHC energies, an interpolation between the two is desirable to clarify the uniqueness of the phenomenon. This is a challenge for the theory, and it can be done within Regge-pole models. 
Below we do so using a simple Regge-pole model with two leading Regge exchanges (the pomeron and odderon) and two secondary reggeons, $f$ and $\omega$. 
%The odderon contribution, however, at low $|t|$ has a small role.    

Having identified \cite{LNC, C-I1, C-I2} the observed "break" with a two-pion exchange effect, 
we investigate further two aspects of the phenomenon, namely: 1) to what extent is the "break" observed recently at the LHC a "recurrence"  of that seen at the ISR (universality)? 2) what is the relative weight of the Regge residue (vertex) compared to the trajectory (propagator) in producing the "break"? We answer these questions by means of a detailed fit to the elastic proton-proton scattering data in the relevant kinematic range $0.01<|t|<0.3$ GeV$^2$ from  the ISR energies ($44.7$ and $52.8$ GeV) up to those at the LHC ($8$ and $13$ TeV).

As shown by Barut and Zwanziger \cite{Barut}, $t$-channel unitarity constrains the Regge trajectories near the threshold, $t\rightarrow t_0$ by
\begin{equation} \label{Eq:Barut}
\Im \, \alpha(t)\sim (t-t_0)^{\Re\, \alpha(t_0)+1/2},
\end{equation} 
where $t_0$ is the lightest threshold, $4m_{\pi}^2$ in the case of the vacuum quantum numbers (pomeron or $f$ meson). Since the asymptotic behavior of the trajectories is constrained by dual models with Mandelstam analyticity by square-root (modulus $\ln t$):
$\mid\frac{\alpha(t)}{\sqrt{t}\ln t}\mid_{t\rightarrow \infty}\leq {\rm const}$, 
(see \cite{LNC} and references therein), for practical reasons it is convenient to approximate, for the region of $t$ in question,  the trajectory as a sum of square roots. This is a simple model satisfying the constraints imposed by analyticity and duality; while the lowest $4m_{\pi}^2$ threshold provides for the imaginary part of the trajectory, heavy thresholds promote the (almost linear) rise of the real part, terminating at the heaviest one. In a limited range, typically that of the diffraction cone, $|t|\le 1.5$ GeV$^2$, for simplicity,  the contribution from heavy thresholds may be approximated by their series expansion, {\it i.e.} by a linear term, as in Eq. (\ref{Eq:trajectory}). Related examples and applications as well as earlier references may be found in Refs. \cite{Biro, Rainer}. We found by trial and error that that the present one -- a two-pion threshold plus a linear term -- is the optimal choice in the kinematic region in question. In paper \cite{Rainer} the almost linear real part of known meson trajectories is related to their imaginary parts by a dispersion relation.

\section{A simple Regge-pole model}\label{Sec:Model}

We use a simple Regge-pole model with leading supercritical pomeron \cite{Landshoff} and odderon and two secondary, $f$ and $\omega$ contributions. The {\bf $\rm pp$} elastic scattering amplitude is:
\begin{equation} \label{Eq:ampl}
A(s,t)=A_P(s,t)-A_O(s,t)+A_f(s,t)-A_\omega(s,t),
\end{equation} 
where
\begin{eqnarray}\label{Eq:Pf}
\begin{aligned}
A_P\left(s,t\right)=a_P{\rm e}^{\beta_P(t)}\Bigl(-is/s_{0}\Bigr)^{\alpha_P\left(t\right)}, \\ 
A_O\left(s,t\right)=ia_O{\rm e}^{b_Ot}\Bigl(-is/s_{0}\Bigr)^{\alpha_O\left(t\right)}, \\
A_f\left(s,t\right)=a_f{\rm e}^{b_ft}\Bigl(-is/s_{0}\Bigr)^{\alpha_f\left(t\right)}, \\ 
A_\omega\left(s,t\right)=ia_\omega{\rm e}^{b_\omega t}\Bigl(-is/s_{0}\Bigr)^{\alpha_\omega\left(t\right)},
\end{aligned}
\end{eqnarray}
with the trajectories
\begin{eqnarray}\label{Eq:trajectory}
\begin{aligned}
\alpha_P\left(t\right)=\alpha_{0P}+\alpha'_Pt-\alpha_{1P}\left(\sqrt{4m_{\pi}^2-t}-2m_{\pi}\right), \\ 
\alpha_O\left(t\right)=\alpha_{0O}+\alpha'_Ot, \\ 
\alpha_f\left(t\right)=0.703+0.84t, \\
\alpha_{\omega}\left(t\right)=0.435+0.93t, \\ 
\end{aligned}
\end{eqnarray}
and
\begin{equation}\label{Eq:residua}
\beta_P(t)=\beta_{0P}+\beta'_Pt-\beta_{1P}\sqrt{4m_{\pi}^2-t}.
\end{equation}
Note that deviation from the linear exponential comes both from the pomeron trajectory, Eq.~(\ref{Eq:trajectory}), and the pomeron residue (vertex), Eq.~(\ref{Eq:residua}), as illustrated schematically in Figure~\ref{Fig:Diagram}~\footnote[1]{For simplicity, here we neglect it in non-leading contributions. Note that in a recent related study \cite{RPM}, the pomeron residue was replaced by $e^{b\alpha(t)}$, where $\alpha(t)$ a non-linear pomeron trajectory.}. Our aim is to answer the questions posed at the end of the the Abstract of the present paper. 

We use the normalization:
\begin{equation}
\sigma_{tot}(s)=\frac{4\pi}{s}\Im A(s,t=0),\  \  \ \frac{d\sigma}{dt}=\frac{\pi}{s^2}|A(s,t)|^2.
\end{equation}
With $s_{0}=1$ GeV$^2$, the model contains 15 free parameters ($a_P$ [$\rm{\sqrt{mbGeV^2}}$], $\beta_{0P}$ [dimensionless], $\beta'_P$ [GeV$^{-2}$], $\beta_{1P}$ \\ {[GeV$^{-1}$]}, $\alpha_{0P}$ [dimensionless], $\alpha'_P$ [GeV$^{-2}$], $\alpha_{1P}$ [GeV$^{-1}$], $a_O$ [$\rm{\sqrt{mbGeV^2}}$], $b_O$ [dimensionless], $\alpha'_O$ [GeV$^{-2}$], $\alpha_{1O}$ [GeV$^{-1}$], $a_f$ [$\rm{\sqrt{mbGeV^2}}$], $b_f$ [dimensionless], $a_\omega$ [$\rm{\sqrt{mbGeV^2}}$], $b_\omega$ [dimensionless]). 
The free parameters of the model were fitted simultaneously to the data on elastic proton-proton differential cross section \cite{TOTEM8,ISR,TOTEM13} in the region of the forward cone, ($t>-0.3$ GeV$^2$) as well on to the data on total cross section \cite{totem7,totem8.3,TOTEM13,PDG,Giani} and $\rho$-parameter \cite{totem8.2,PDG,TOTEM13} in the energy range $20\div57000$ GeV by using Minuit. The $\rho$-parameter is the ratio of real and imaginary part of the forward scattering amplitude: 
\begin{equation}
\rho(s)=\frac{\Re A(s,t=0)}{\Im A(s,t=0)}.
\end{equation}

\begin{figure}
	\centering
	\includegraphics[width=1\columnwidth]{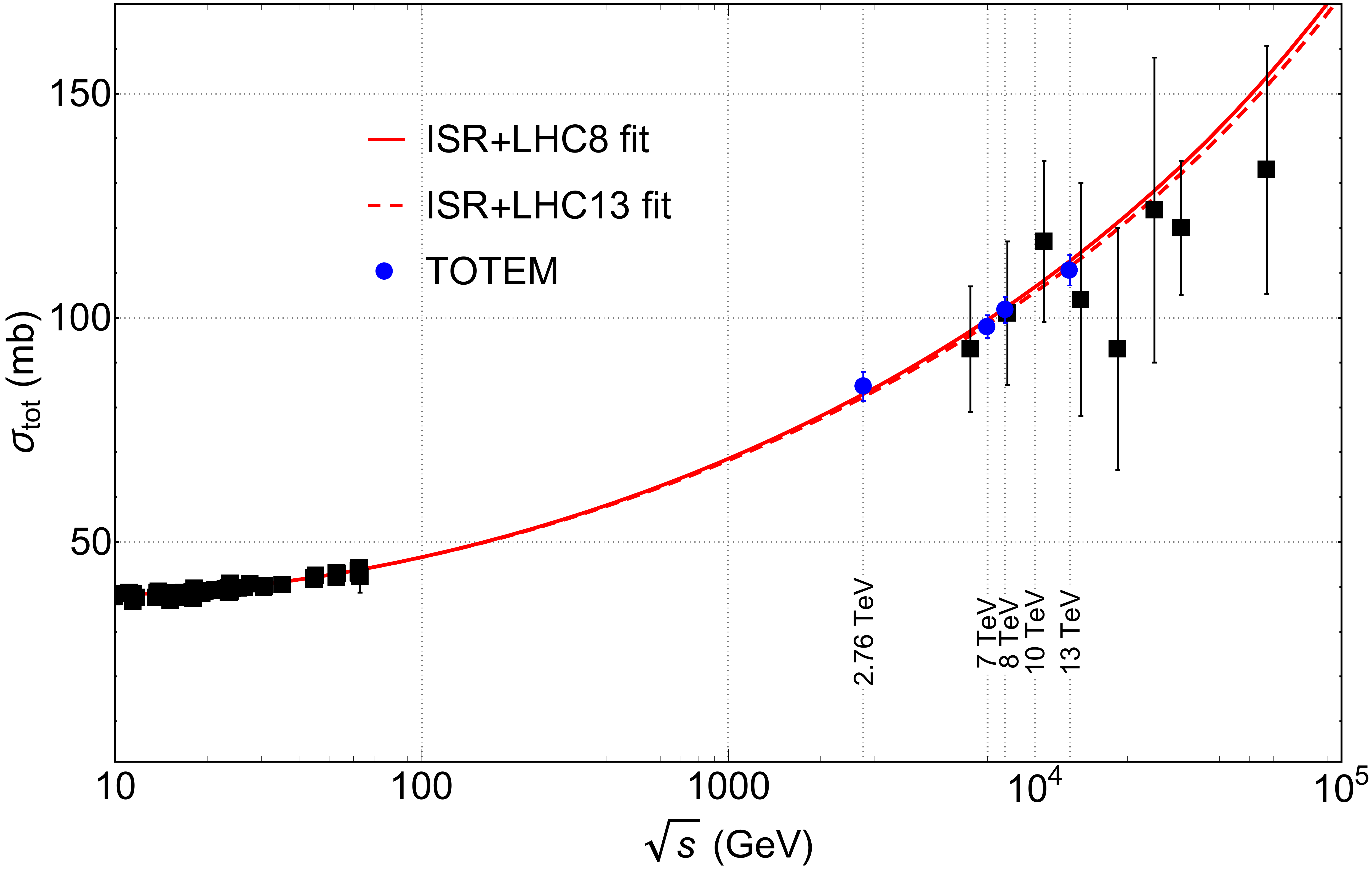}
	\hfill
	\includegraphics[width=1\columnwidth]{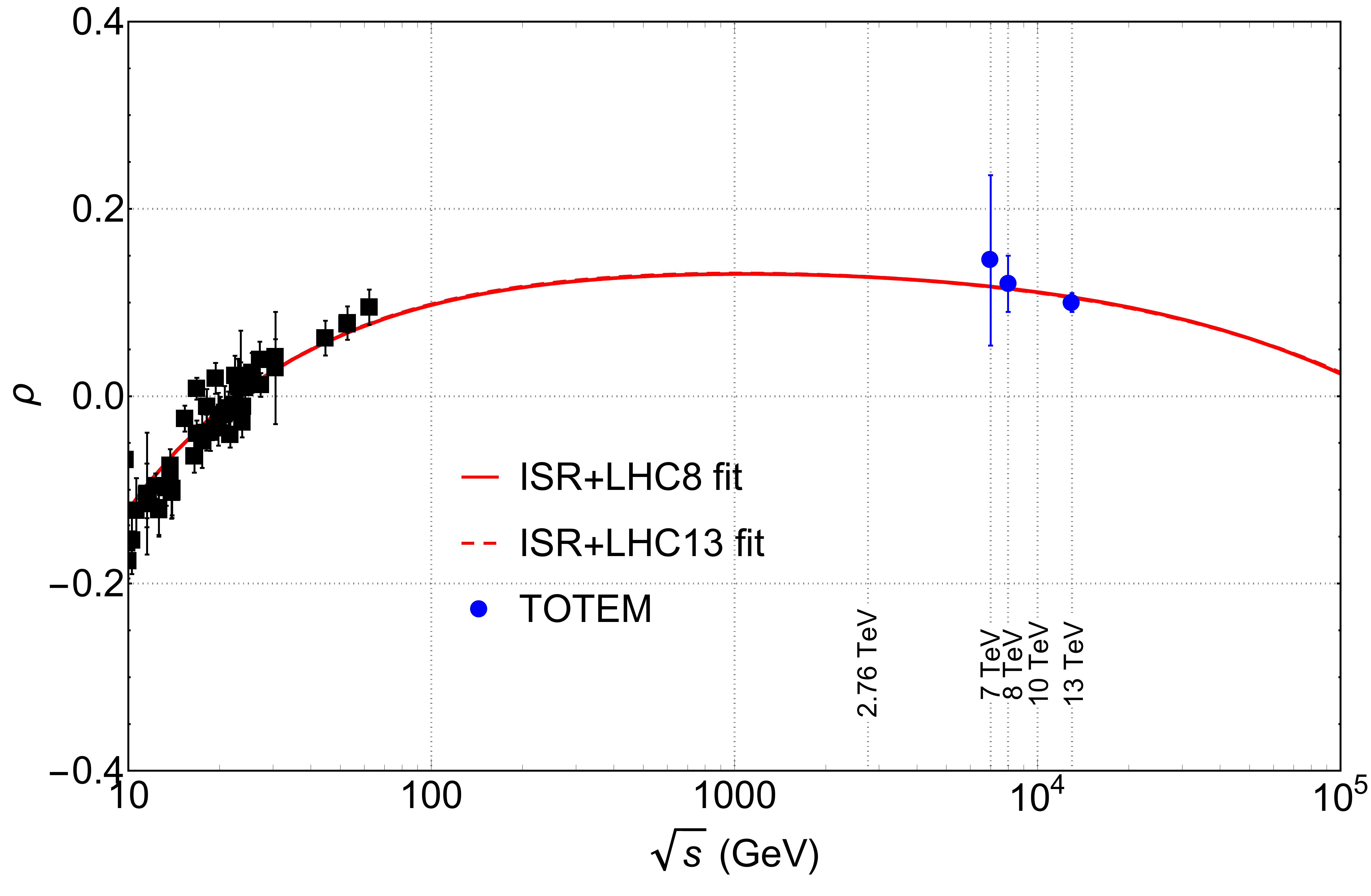}
	\hfill
    \includegraphics[width=1\columnwidth]{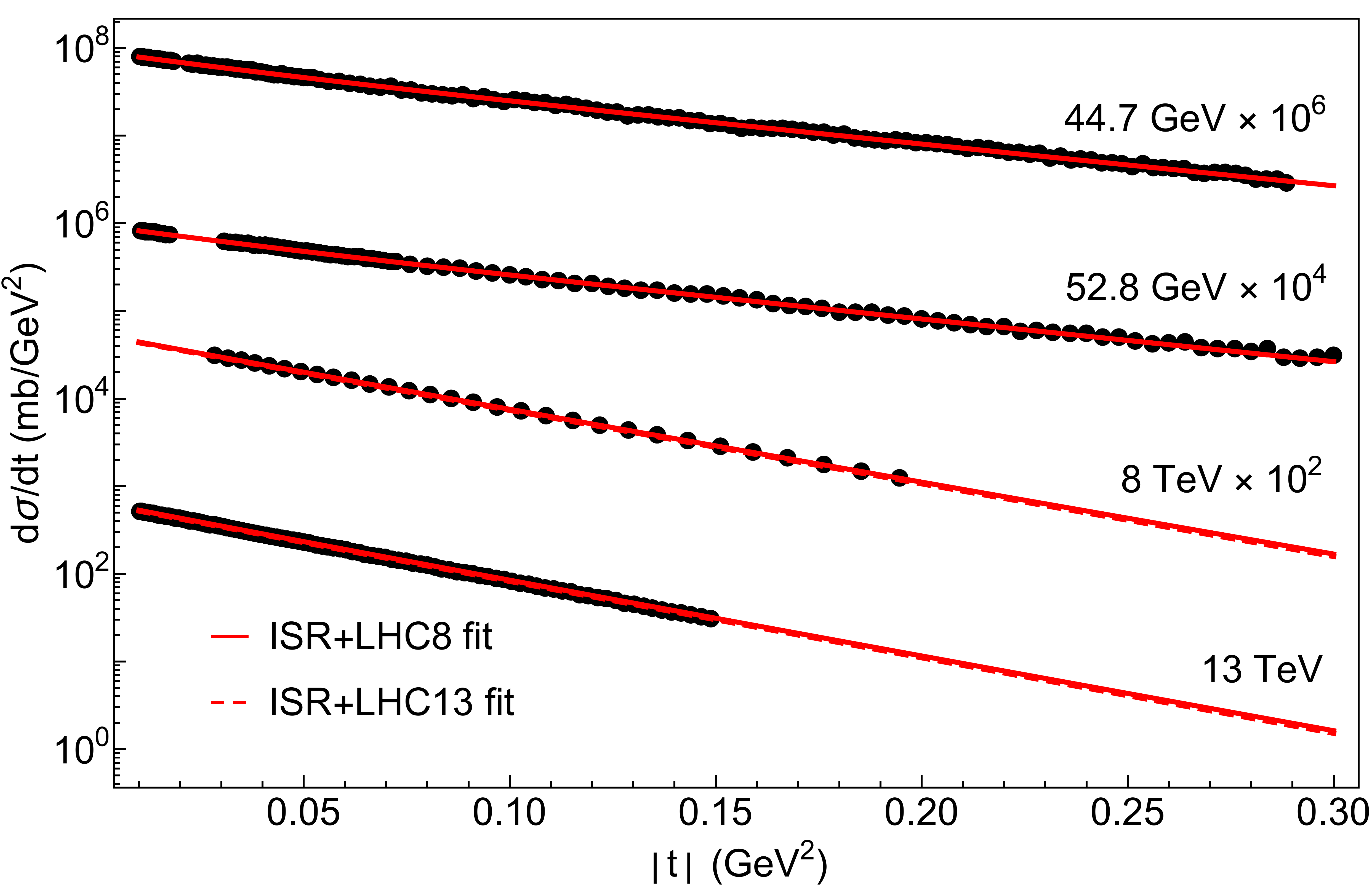}
	\caption{Fit to $pp$ total cross section (top), $\rho$-parameter (middle) and low-$|t|$ differential cross section data (bottom).}
	\label{Fig:dsigma}
\end{figure}

We have investigated several options using a linear and non-linear pomeron trajectory, varying the fitted $s$ and $t$ ranges, The best fits are shown in Figure~\ref{Fig:dsigma} with the values of the fitted parameters quoted in Table~\ref{tab:parameters}. Aiming at a better fit, the number of the ISR data points on the differential cross section was restricted to the chosen $t$ interval, furthermore the LHC 13 TeV data were fitted up to $|t_{max}|=0.15$, as it was done by TOTEM in Ref.~\cite{TOTEM13}.  

Examination of the relative role/weight of the (non-exponential) residue compared to the (non-linear) pomeron trajectory is among the main objectives of the present work, aimed at a better understanding of the matter distribution in nuclei.

\begin{table}
	\caption{Values of the parameters fitted to the $pp$  data on the $\rho$-parameter, total and differential cross section.}
	\centering
	\subfloat[ISR + LHC 8 TeV \label{sfig:testa}]{%
		\begin{tabular*}{\columnwidth}{@{\extracolsep{\fill}}lllll@{}}\hline
		&	Parameter& Value & Error \\\hline
		&	$a_P$ &-1.62206&0.00723783&\\
		&	$\alpha_{0P}$& 1.09505&0.000507519&\\
		&	$\alpha'_P$& 0.350352&0.000807463&\\
		&	$\alpha_{1P}$&0.0418504&0.000894104&\\
		&	$\beta_{0P}$& 0.825955&0.00283623&\\
		&	$\beta'_P$&2.52918&0.0290946&\\
		&	$\beta_{1P}$&-0.036672&0.00831786&\\
		&	$a_O$&0.00113782&0.000135669&\\
		&	$b_O$&2&fixed&\\
		&	$\alpha_{0O}$&1.36284&0.00461602&\\
		&	$\alpha'_O$& 0.4&fixed&\\
		&	$a_f$&-11.6528&0.257487&\\
		&	$b_f$&13.8938&0.868578&\\
		&	$a_\omega$&9.92422&1.14544&\\
		&	$b_\omega$&10&fixed&\\
		&	$s_{0}$&1 & fixed& \\\hline \hline
		&	&$\chi^2/DOF$ &1.3&\\
		&	&$DOF$ & 183 &\\\hline
		\end{tabular*}%
	}\qquad
	\subfloat[ISR + LHC 13 TeV\label{sfig:testb}]{%
		\begin{tabular*}{\columnwidth}{@{\extracolsep{\fill}}lllll@{}}\hline
		&	Parameter& Value & Error \\\hline
&	$a_P$ &-1.63005&0.00719853&\\
&	$\alpha_{0P}$& 1.09385&0.000548782&\\
&	$\alpha'_P$& 0.361809&0.00149993&\\
&	$\alpha_{1P}$&0.0372772&0.000761573&\\
&	$\beta_{0P}$&0.832661&0.00282508&\\
&	$\beta'_P$&2.49077&0.0282044&\\
&	$\beta_{1P}$&-0.0364331&0.00852847&\\
&	$a_O$&0.000860881&0.000126785&\\
&	$b_O$&2&fixed&\\
&	$\alpha_{0O}$&1.37452&0.00331293&\\
&	$\alpha'_O$& 0.4&fixed&\\
&	$a_f$&-11.486&0.259798&\\
&	$b_f$&14.3807&0.907834&\\
&	$a_\omega$&10.1825&1.12135&\\
&	$b_\omega$&10&fixed&\\
&	$s_{0}$&1 & fixed& \\\hline \hline
&	&$\chi^2/DOF$ &1.2&\\
&	&$DOF$ & 246 &\\\hline 
		\end{tabular*}%
	}     
	\label{tab:parameters}
\end{table}

\section{Mapping the "low-energy" break to that at the LHC} \label{Extrapolate}

\begin{figure*}
	\centering
	\includegraphics[width=1\columnwidth]{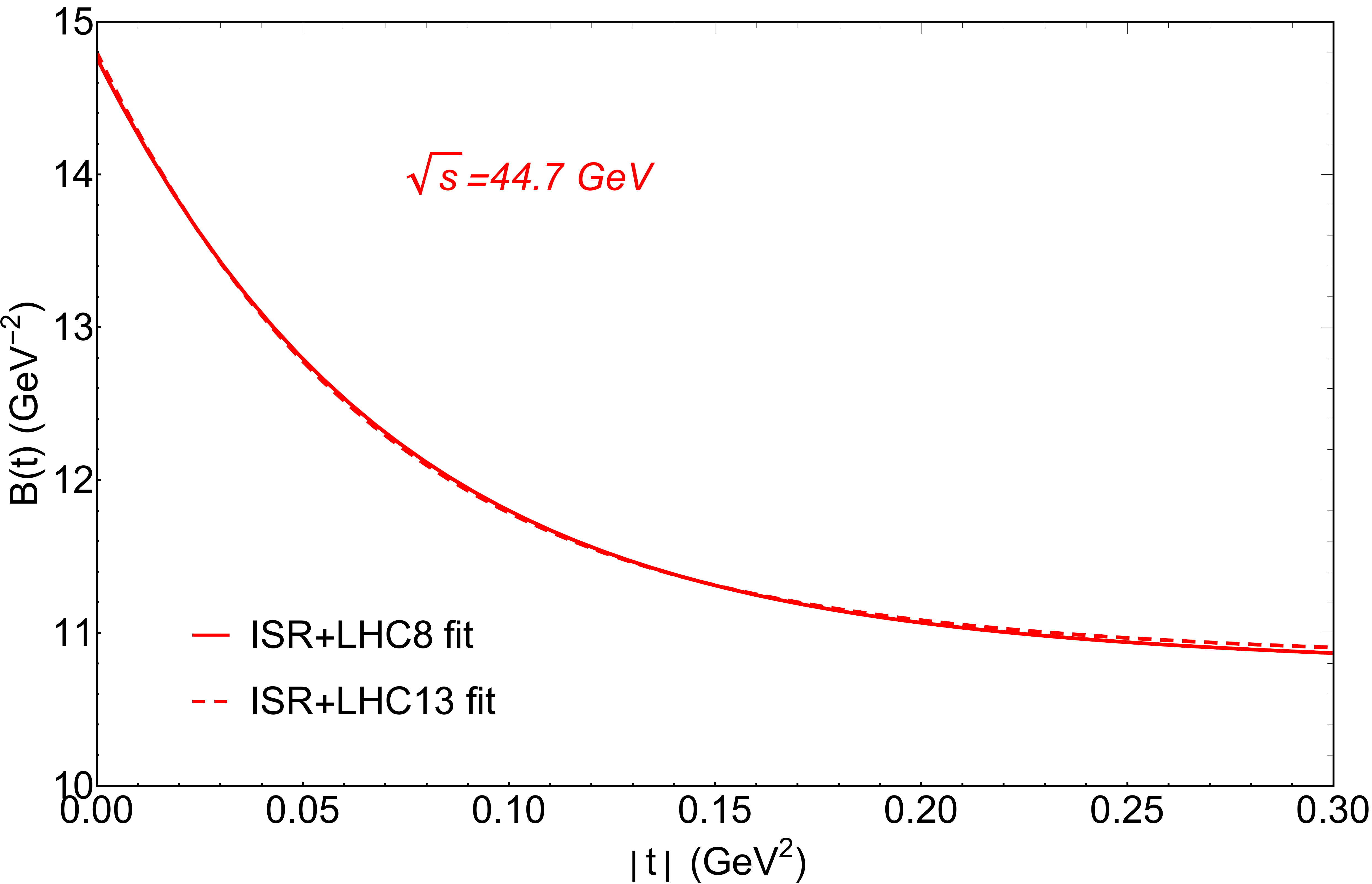}
	\hfill
	\includegraphics[width=1\columnwidth]{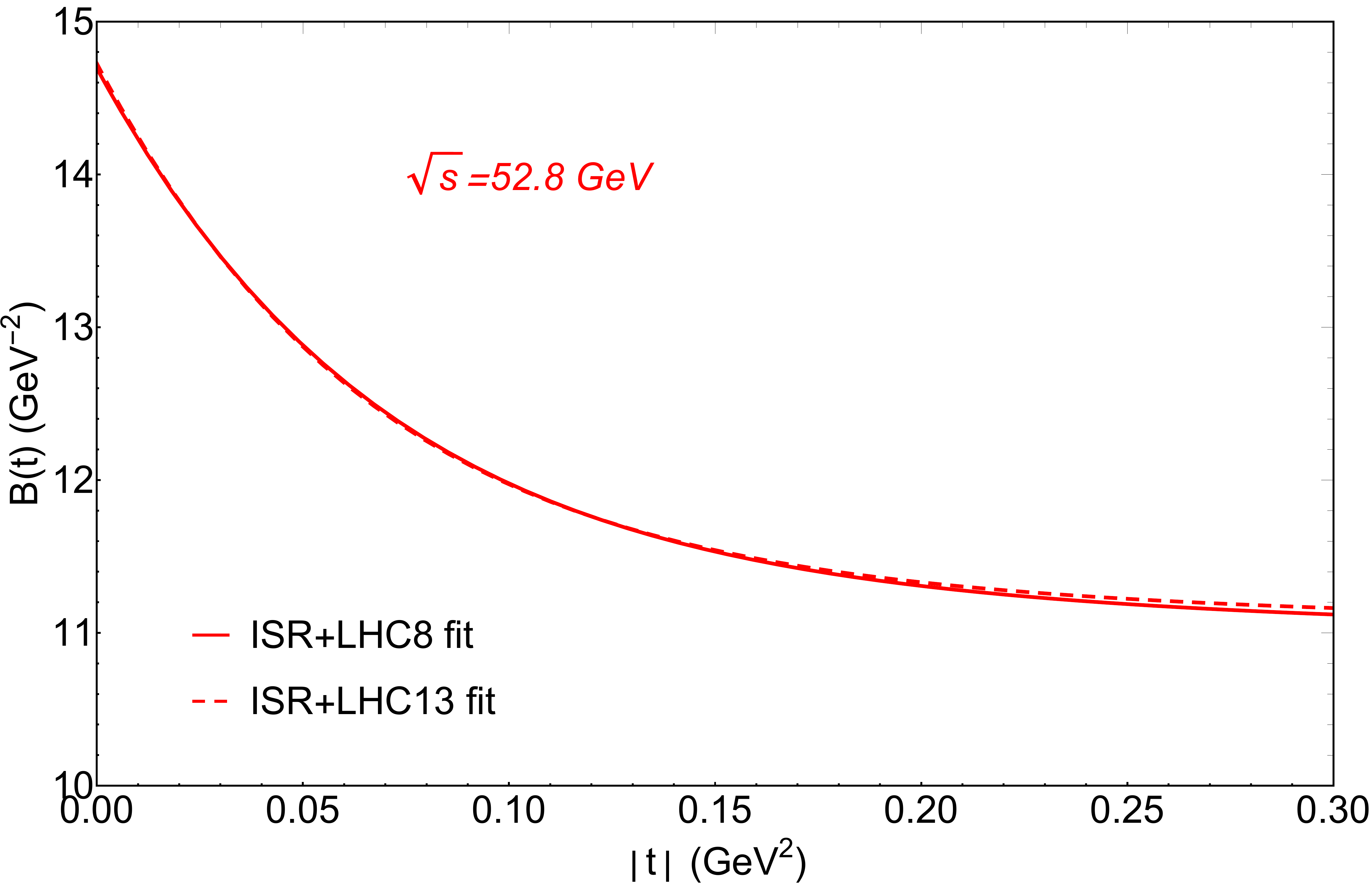}
	\hfill
    \includegraphics[width=1\columnwidth]{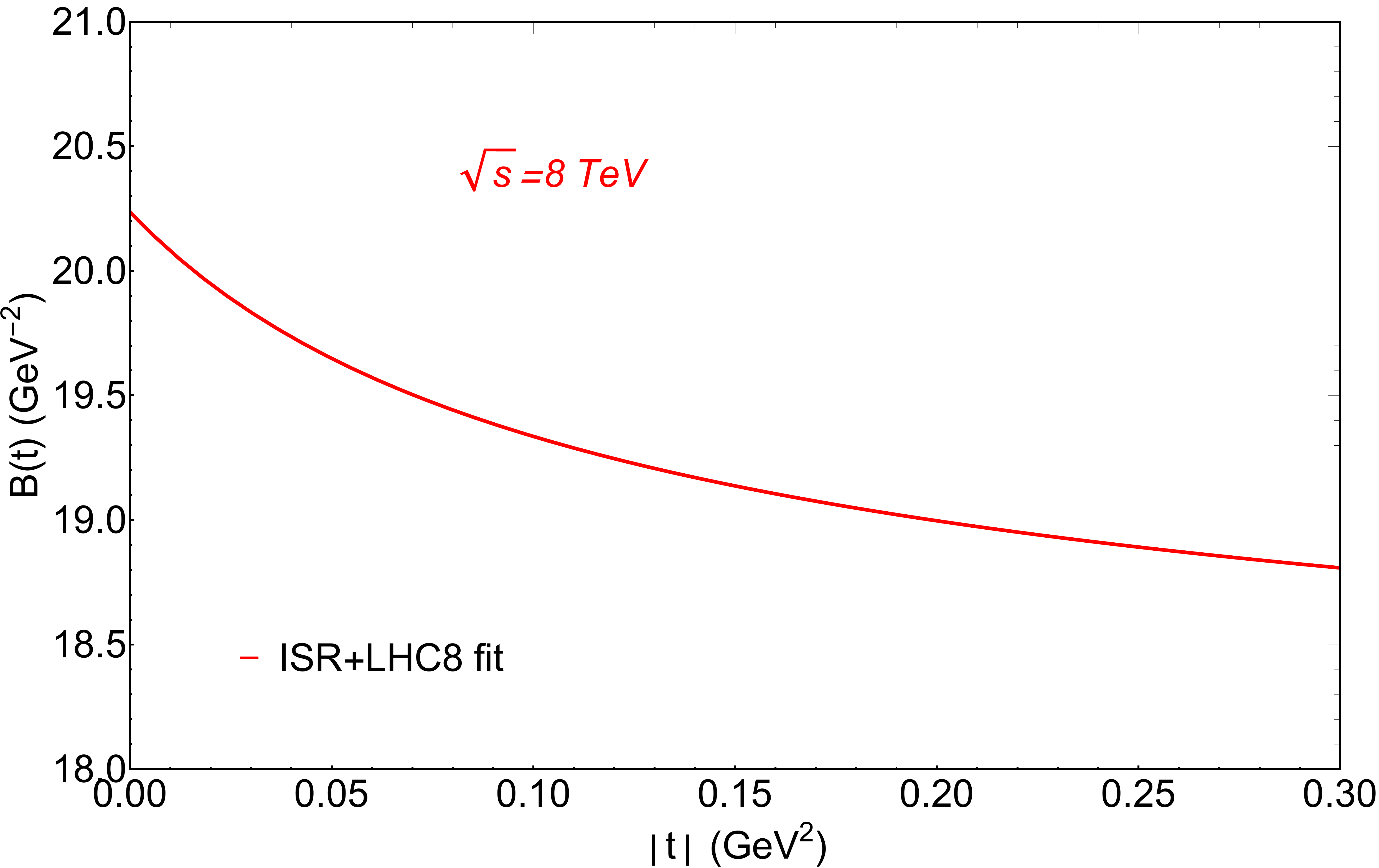}
	\hfill
	\includegraphics[width=1\columnwidth]{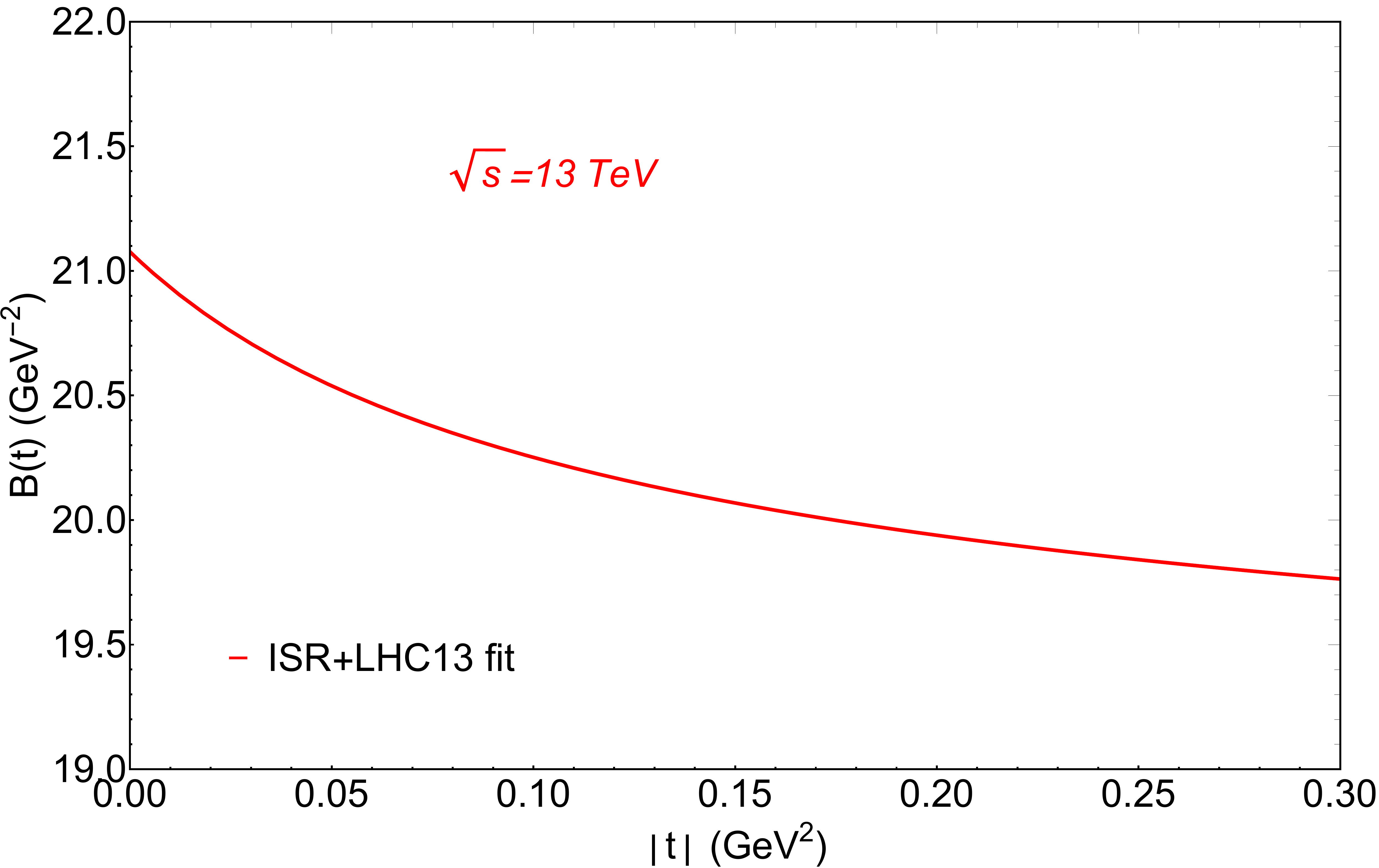}
	\caption{Local slopes.}
	\label{Fig:localslope}
\end{figure*}

\begin{figure*}
	\centering
	\includegraphics[width=1\columnwidth]{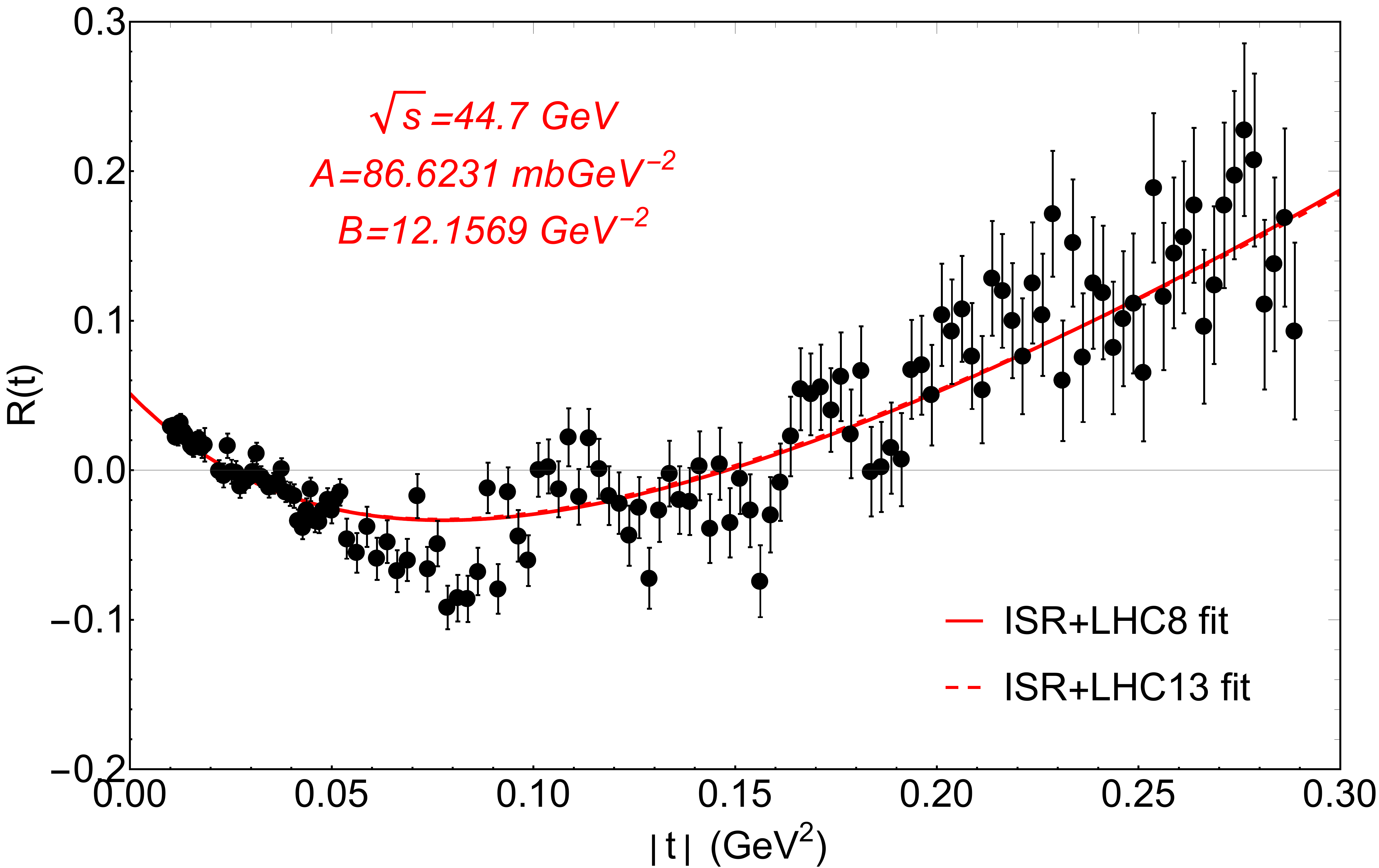}
	\hfill
	\includegraphics[width=1\columnwidth]{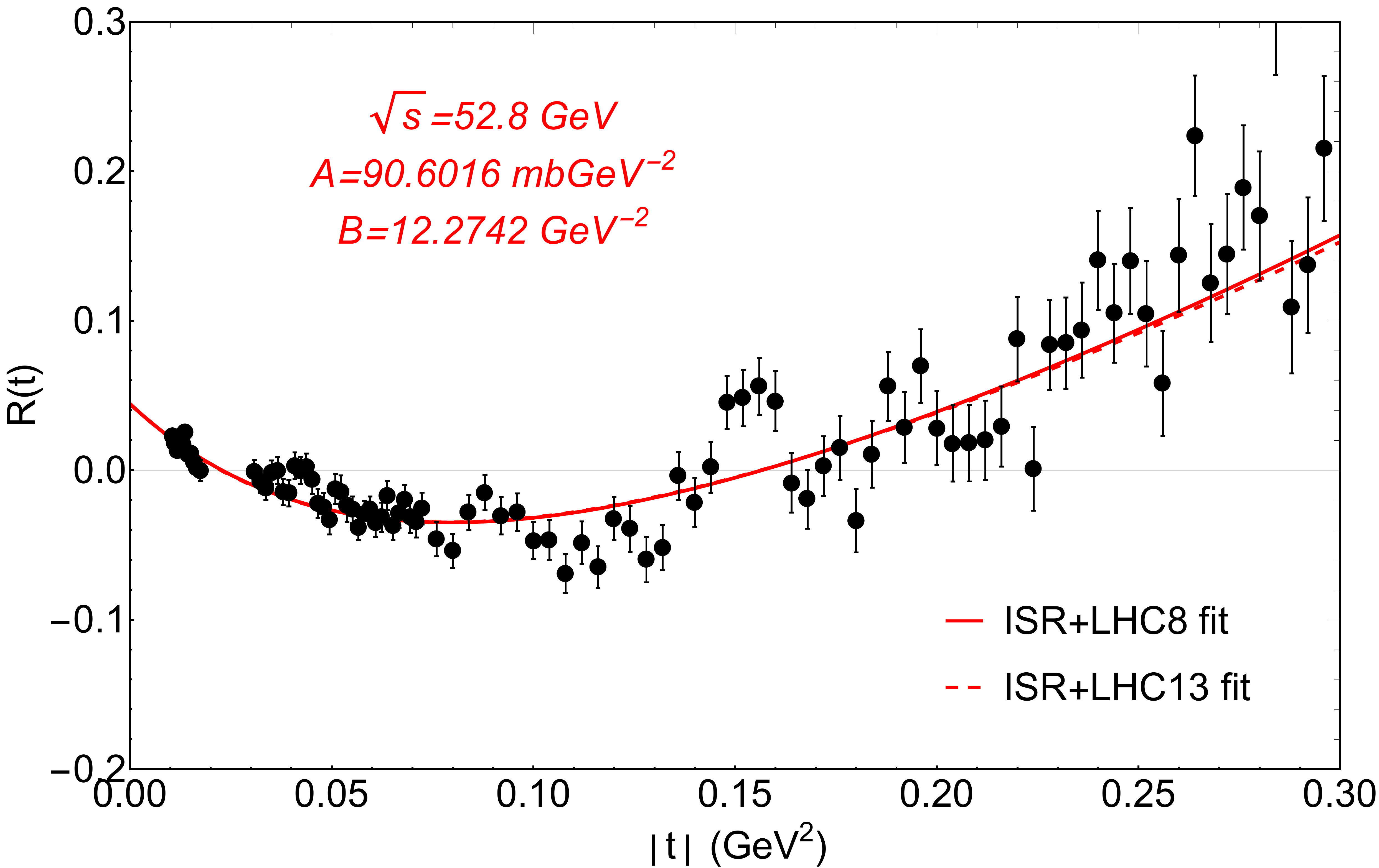}
	\hfill
	\includegraphics[width=1\columnwidth]{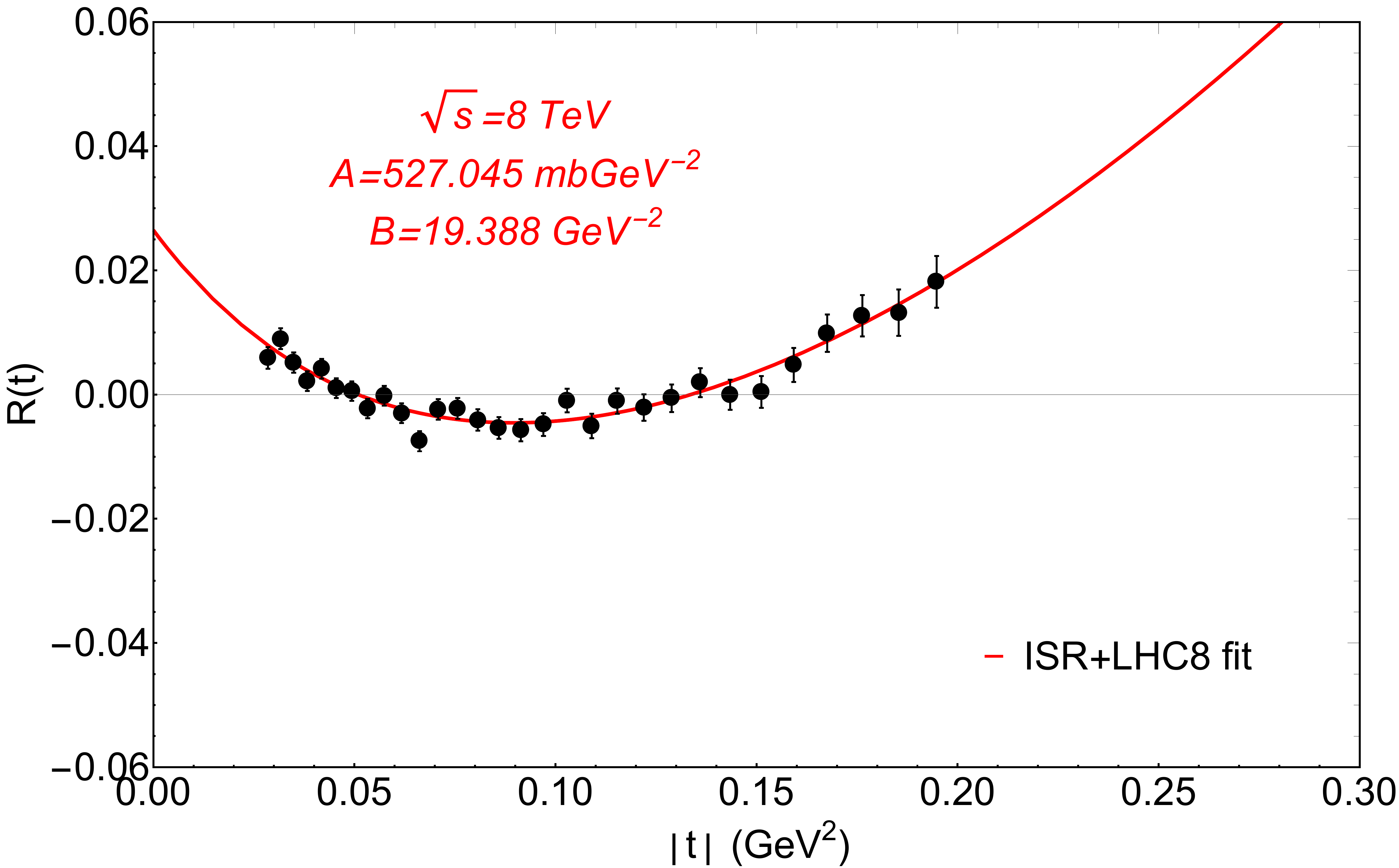}
	\hfill
	\includegraphics[width=1\columnwidth]{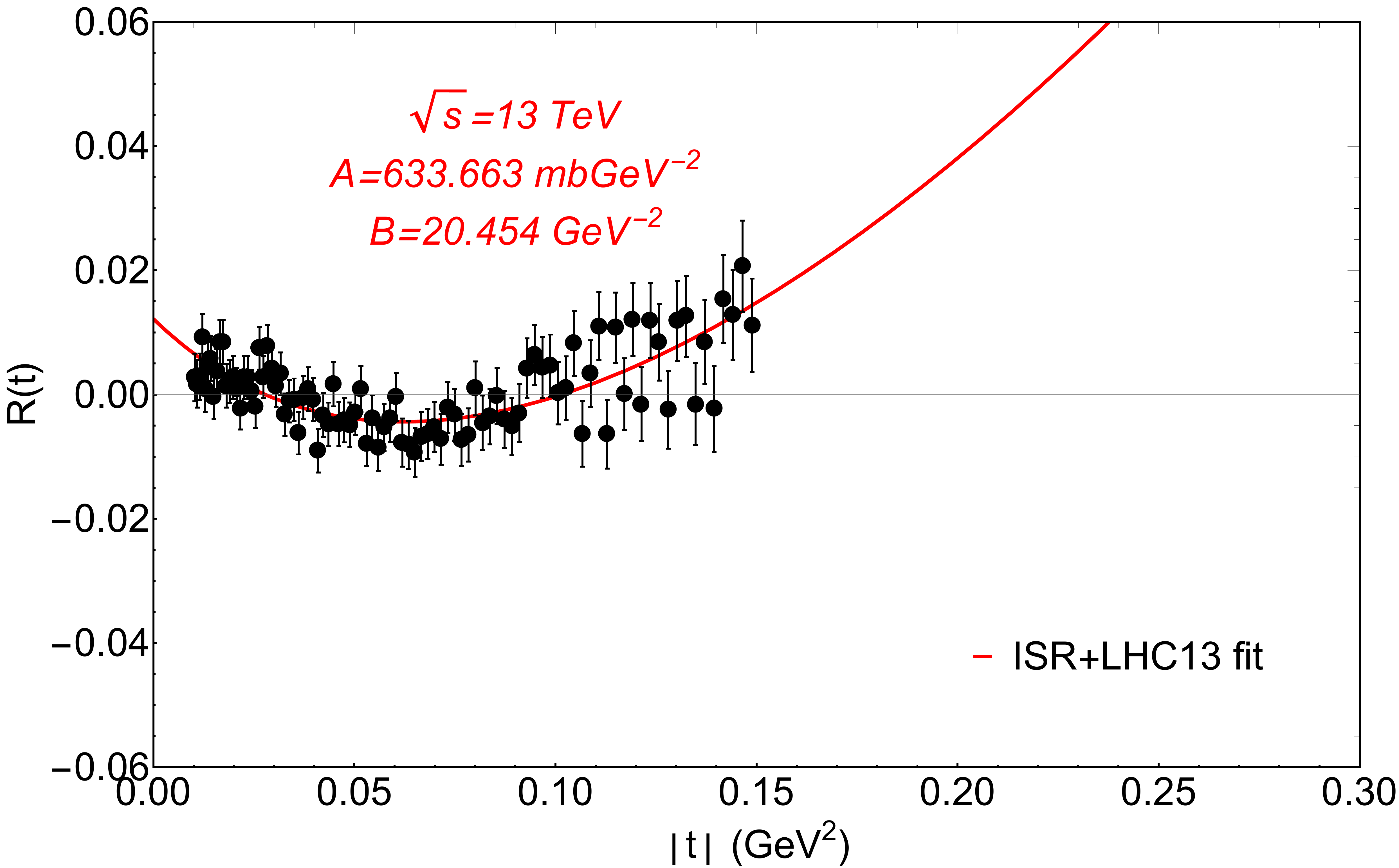}
	\caption{$R(t)$ ratios.}
	\label{Fig:Rratio}
\end{figure*}

\begin{table*}
	\caption{Correlation matrix in case of a fit to the ISR + LHC (8 TeV)  data.}
	\centering
	\label{cor8}
	\begin{tabular}{|c|c|c|c|c|c|c|c|c|c|c|c|c|}
		\hline
		&$a_P$&$\alpha_{0P}$&$\beta_{0P}$&$a_f$&$a_\omega$&$a_O$&$\alpha_{0O}$&$\alpha'_P$&$\alpha_{1P}$&$\beta'_{P}$&$\beta_{1P}$&$b_f$\\ \hline
		$a_P$&1&-0.89&0.996&0.802&0.311&0.565&0.241&-0.246&-0.811&0.57&0.363&0.502 \\
		$\alpha_{0P}$&-0.89&1&-0.85&-0.662&-0.141&-0.835&-0.561&-0.0377&0.77&-0.527&-0.0401&-0.521 \\
		$\beta_{0P}$&0.996&-0.859&1&0.805&0.33&0.521&0.199&-0.276&-0.797&0.556&0.41&0.487\\
		$a_f$&0.802&-0.662&0.805&1&-0.182&0.321&0.0279&-0.326&-0.691&0.549&0.272&0.428\\
		$a_\omega$&0.311&-0.141&0.33&-0.182&1&-0.0195&-0.111&-0.212&-0.189&0.107&0.348&0.0342\\
		$a_O$&0.565&-0.835&0.521&0.321&-0.0195&1&0.921&0.406&-0.436&0.157&0.00527&0.157\\
		$\alpha_{0O}$&0.241&-0.561&0.199&0.0279&-0.111&0.921&1&0.552&-0.138&-0.12&0.0186&-0.159\\
		$\alpha'_P$&-0.246&-0.0377&-0.276&-0.326&-0.212&0.406&0.552&1&0.585&-0.715&-0.3&0.0215\\
		$\alpha_{1P}$&-0.811&0.77&-0.797&-0.691&-0.189&-0.436&-0.138&0.585&1&-0.916&-0.112&-0.395\\
		$\beta'_{P}$&0.57&-0.527&0.556&0.549&0.107&0.157&-0.12&-0.715&-0.916&1&-0.12&0.40\\
		$\beta_{1P}$&0.363&-0.0401&0.41&0.272&0.348&0.00527&0.0186&-0.3&-0.112&-0.12&1&-0.401\\
		$b_f$&0.502&-0.521&0.487&0.428&0.0342&0.157&-0.159&0.0215&-0.395&0.4&-0.401&1\\ \hline
	\end{tabular}
\end{table*}

\begin{table*}
	\caption{Correlation matrix in case of a fit to the ISR + LHC (13 TeV) data.}
	\centering
	\label{cor13}
	\begin{tabular}{|c|c|c|c|c|c|c|c|c|c|c|c|c|}
		\hline
		&$a_P$&$\alpha_{0P}$&$\beta_{0P}$&$a_f$&$a_\omega$&$a_O$&$\alpha_{0O}$&$\alpha'_P$&$\alpha_{1P}$&$\beta'_{P}$&$\beta_{1P}$&$b_f$\\ \hline
		$a_P$& 1&-0.874&0.995&0.79&0.257&0.549&0.522&-0.92&0.446&0.575&0.28&0.461 \\
		$\alpha_{0P}$&-0.874&1&-0.845&-0.609&-0.0775&-0.846&-0.825&0.81&-0.66&-0.494&0.0689&-0.473 \\
		$\beta_{0P}$&0.995&-0.845&1&0.796&0.274&0.51&0.484&-0.914&0.42&0.565&0.323&0.448\\
		$a_f$&0.790&-0.609&0.796&1&-0.222&0.261&0.2292&-0.748&0.216&0.546&0.232&0.402\\
		$a_\omega$&0.257&-0.0775&0.274&-0.222&1&-0.0625&-0.0637&-0.218&-0.0487&0.109&0.329&-0.021\\
		$a_O$&0.549&-0.846&0.51&0.261&-0.0625&1&0.996&-0.422&0.692&0.123&-0.088&0.105\\
		$\alpha_{0O}$&0.522&-0.825&0.484&0.229&-0.0637&0.996&1&-0.388&0.7&0.0806&-0.0799&0.0882\\
		$\alpha'_P$&-0.920&0.81&-0.914&-0.748&-0.218&-0.422&-0.388&1&-0.233&-0.814&-0.0984&-0.559\\
		$\alpha_{1P}$&0.446&-0.66&0.42&0.216&-0.0487&0.692&0.7&-0.233&1&-0.202&-0.227&0.5\\
		$\beta'_{P}$&0.575&-0.494&0.565&0.546&0.109&0.123&0.0806&-0.814&-0.202&1&-0.161&0.432\\
		$\beta_{1P}$&0.280&0.0689&0.323&0.232&0.329&-0.088&-0.0799&-0.0984&-0.227&-0.161&1&-0.482\\
		$b_f$&0.461&-0.473&0.448&0.402&-0.0218&0.105&0.0882&-0.559&0.5&0.432&-0.482&1\\ \hline
	\end{tabular}
\end{table*}

At the ISR, the proton-proton differential cross section was measured in the energy range $23.5\leq\sqrt{s}\leq62.5$ GeV \cite{ISR}. At those energies the diffraction cone changes its slope near $-t=0.1$ GeV$^2$ by about $2$ GeV$^{-2}$. Below we use the Regge-pole model introduced in the previous section to map this "break", fitted to the ISR data onto the LHC TOTEM 8 and 13 TeV data. The result is shown in Figure~\ref{Fig:dsigma}. To highlight this phenomenon, we show it in details and better resolution in Figs. ~\ref{Fig:localslope} and~\ref{Fig:Rratio}. In Figure~\ref{Fig:localslope}, we show the local slopes
\begin{equation}
B(s,t)=\frac{d}{dt} \ln (d\sigma/dt ) \,,
\end{equation}
resulting from for our fits at four values of $s$. To demonstrate the quality of our fit and to anticipate the comparison with the LHC data, we  present  in Figure~\ref{Fig:Rratio}  the ISR data also in a normalized form used by TOTEM \cite{TOTEM8}: 
\begin{equation} \label{Eq:norm}
R=\frac{d\sigma/dt}{d\sigma/dt_{ref}}-1,
\end{equation}
where $d\sigma/dt_{ref}=Ae^{Bt}$, with $A$ and $B$ constants determined from a fit to the experimental data.   

Both  Figure~\ref{Fig:localslope} and Figure~\ref{Fig:Rratio} re-confirm the earlier finding that the "break" can be attributed the presence of two-pion branch cuts in the Regge parametrization.

To summarize, the best description of the data is obtained when both the  pomeron trajectory is non-linear and the residue is non-exponential. The fit quality deteriorates if the pomeron trajectory is set linear or the residue is exponential i.e. $\alpha_{1P}$ or $\beta_{1P}$ are set zero. The correlation matrices obtained from the fits are presented in Table~\ref{cor8} and Table~\ref{cor13}. The correlation between the mentioned parameters is not large: by setting one of them zero the fit deteriorates: in case of the ISR+LHC8  data set, $\chi^2/DOF\sim$2.1, while in case of ISR+LHC13 $\chi^2/DOF\sim 1.6$. It can be concluded from these results that, concerning the "break", the Regge residue and the pomeron trajectory have nearly the same weight and importance.

\section{Size and Shape of Proton}

In a geometric approach to diffraction scattering, both the total cross section and the elastic slope $B$ are directly related to the size and shape of the hadrons. However, from a QCD perspective, these concepts do not enter directly, and they should emerge from first principle, albeit non-perturbatively.   For a proton at rest, the concept of ``static size", measured by ``weak probe",  can be defined, i.e., cross sections in the  ``low-energy limit".  In particular, the scale for the  static size should reflect dominant QCD dynamics at low energy. It is generally accepted that  low-energy QCD dynamics is controlled by chiral symmetry breaking involving quark degrees of freedom, e.g., leading to constituent quark masses.  In working with chiral lagrangian, pion mass and coupling play a central role.

In the high energy limit, however, due to Lorentz contraction, each appears as a "disk". Due to vacuum fluctuations,  a proton begins to reveal its partonic structure, leading to an increase in its transverse size, i.e., increasing total cross section.   The process is driven by inelastic production through a diffusion mechanism, with~\footnote[2]{In an eikonal approach, this also leads to $\sigma_{total}\simeq b \Delta \alpha_0\pi  \ln^2 s +{\rm subleading \, terms}$, where $\Delta \alpha_0=\alpha_0-1$.}
\begin{equation}\label{slope}
B(s,0)\simeq b \ln(s/s_0) + c\, .
\end{equation}
Since the process is non-perturbative, it is reasonable that the coefficient $b$  is of confinement scale, e.g., ${\rm GeV}^{-2}$. In contrast,  coefficient $c$,  which dominates in the low-energy limit, can be sensitive to pion mass. For example, for our preferred model with linear pomeron trajectory, $b\simeq 2\alpha'_P$ and $c\simeq \beta'_P+ \beta_{1P}/(4 m_{\pi})$. In particular, $m_{\pi}^{-1}$  sets the scale for the low-energy proton size, i.e., the associated pion cloud, surrounding the constituent quarks of a proton.  
This physically motivated  picture in interpreting elastic peak and hadron size was explained more fully in \cite{C-I1, C-I2}. 

\begin{figure}
	\centering
	\includegraphics[width=1\columnwidth]{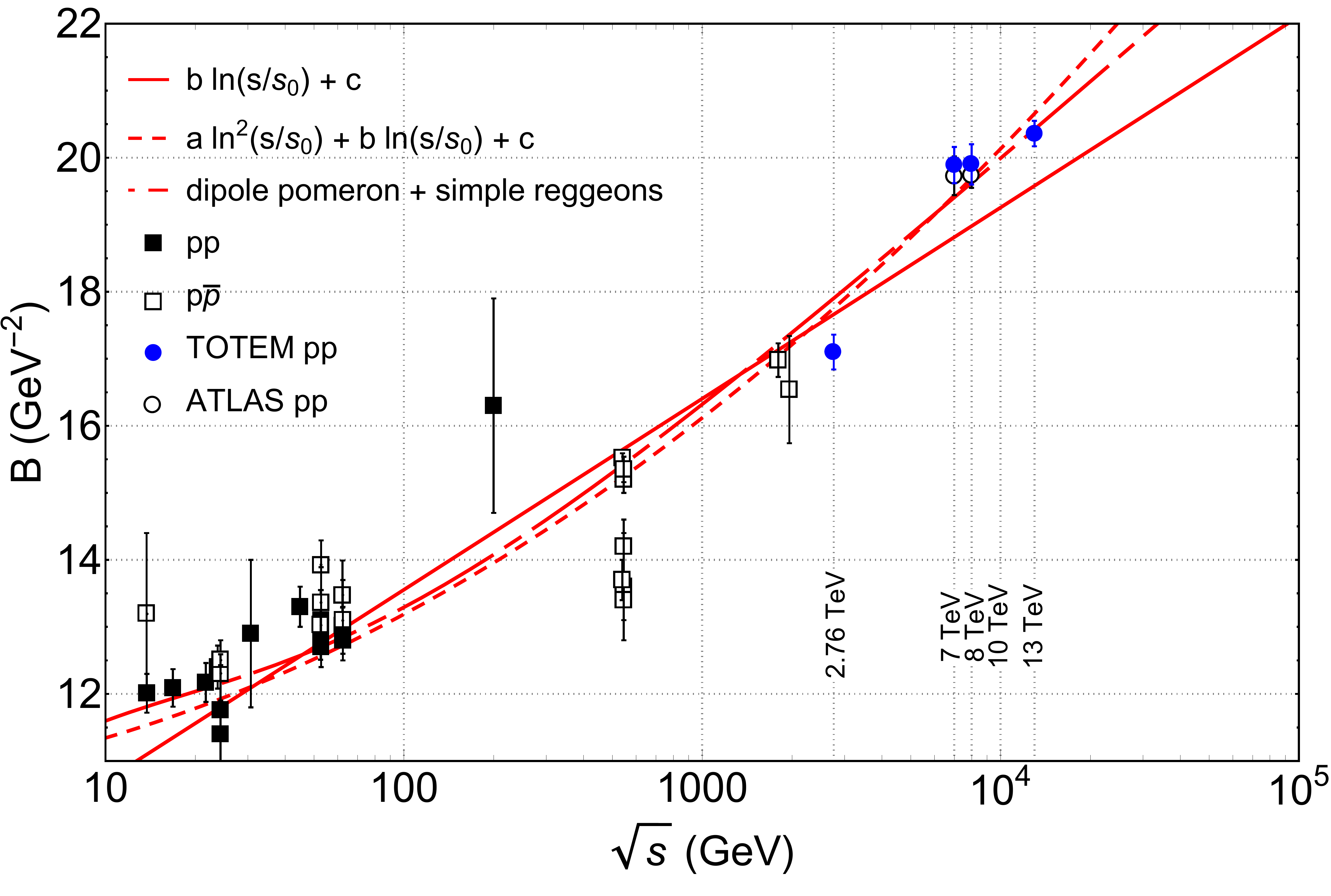}
	\caption{Fits \cite{BJSz} to the $pp$ and $p \bar p$ elastic slope data \cite{ISR,Giani,atlas7,atlas8,totem7,totem8.3}.}
	\label{Fig:slope}
\end{figure}

A simple fit to the Regge-pole-motivated logarithmic slope Eq.~(\ref{slope}) offers the following values of the parameters: $b=0.6191,\ \ c=7.852$ in unites of GeV$^{-2}$. However, recent TOTEM measurements \cite{Giani} show a drastic deviation (acceleration) from the logarithmic to $\ln^2 s$ rise of $B(s)$ beyond $\sqrt{s}\approx3$ TeV, anticipated in \cite{Sch}. With the $\ln^2(s)$ term included, $B(s)\sim a\ln^2(s)+b\ln(s)+c$, the fit to the data, including those recent from TOTEM give $a=0.02557$, $b=0.04715$, $c=10.58$ although the fit does not seem definite. Figure~\ref{Fig:slope}
shows fits with a dipole pomeron (see \cite{BJSz} and earlier references therein) appended by secondary, $f$ and $\omega$ reggeons. In any case, more data points on $B(s)$ at other LHC energies are needed to clarify the deviation of the slope from the expected logarithmic rise (relevant experiments are planned). On the theoretical side, the deviation from the logarithmic rise of $B(s)$, predicted by the unitarized Regge-pole models and by the diffusion mechanism \cite{C-I1, C-I2, Sch} may indicate the importance of unitarity corrections to the Regge-pole amplitude \cite{BJSz}. 

In this study we avoided the dip-bump region, which is an important and complicated issue by itself. Here we only mention that despite intense work along these lines (see e.g. \cite{JLL} and earlier references therein), the existing models or theories are not able to predict unambiguously the details of this important phenomenon. Most of those based on multiple scattering or eikonal formalism predict a sequence of structures, contradicting the experimental data. At the dip, unlike the forward region, the odderon may become visible, moreover important, making different its shape in $pp$ and $\bar pp$ scattering, however increasing also the number of degrees of freedom. New data from the TOTEM Collaboration at the LHC at $13$ TeV will become public soon.

Although we have carried out our phenomenology analysis in terms of 
a Regge model,  the importance of the "static pion cloud" for proton-proton elastic peak can also be framed in a broader context.   The notion of  "elastic peak as the shadow of multiparticle production" has been  a well-established principle, providing a broader link between elastic scattering and inelastic production \cite{Chew1, Chew2, Amati1, Amati2, VanHove}. In this context,   the physically motivated picture of "pion dominance" for regge residue can be extended to the study of  particle dynamics in particle production. This will most likely manifest itself  in the spectrum for the  production of the leading particle at high energy  \cite{C-I2}. In the so-called triple-Regge region,  this enters through the $\pi$-$\pi$-Pomeron triple-Regge contribution.   In particular, in a proton initiated production process, due to proton-neutron-pion coupling, the rate of leading neutron production  should be large and therefore cannot be ignored.  This should have significant impact on understanding leading particle spectrum in air-shower in cosmic ray physics and can be tested at LHC by their combined analysis, see \cite{Cosmic}.

\section{Conclusions} \label{Sec:Conclude}

We have shown that the deviation from a linear exponential of the $pp$ diffraction cone as seen at the ISR, $23.5\leq\sqrt{s}\leq 62.5$ GeV and at the LHC, $\sqrt s=8$ and $13$ TeV are of similar nature: they appear nearly at the same value of $t\approx -0.1$~GeV$^2$, have the same shape of comparable size, $\Delta B(t)\approx 2\div 3$ GeV$^{-2}$ and may be fitted by similar $t$-dependent function. Mapping this $t$-dependence through the tremendous energy span from the ISR to the LHC (almost 3 orders of magnitude) is a highly non-trivial task. Within a simple Regge-pole model with a pomeron and odderon exchange appended with two sub-leading  reggeons, $f$ and $\omega$ we have estimated the relative contribution to the "break" from the non-exponential residue and non-linear pomeron trajectory.

Our findings call also for further studies, namely:

1) theoretical calculations of the relative weight of the loop contribution, second term in Figure~\ref{Fig:Diagram} relative to the first one ("Born term") are needed; 

2) since the pomeron is universal, the effect is expected also in $p\bar p$. Non-observation of a similar structure in the diffraction cone at the Tevatron may be attributed to poor statistics of the relevant data (lacking Roman pots on both sides), preventing the observation of such a tiny effect.

3) the "break" under discussion must be present also in single-, double- and central diffraction dissociation, however to be seen, better statistics on $t$ dependence is needed;

4) the "break" should be present and can be predicted also at still higher energies, however in any approximation beyond the LHC energy, the dip (located near $-t\approx 0.5$ GeV$^2$) will come close to the "break", thus distorting it \cite{JSZ2}. The fate of the diffraction cone at super-high energies (at the Future Circular Collider (FCC)?) is an interesting open question.

To conclude, we expect more precise data in the low-$|t|$ region on elastic scattering and diffraction dissociation as well as further fits with improved phenomenological paramet-rizations. Theoretical calculations of the diagram (Figure~\ref{Fig:Diagram}) may shed more light on the nature of the phenomenon. Further studies in this direction will be based on improved models for the scattering amplitude, with more details on individual Regge trajectories.

\section*{Acknowledgments}
We thank Rainer Schicker and the Heraeus Physics School on QCD (Bad-Honnef, Germany, 2017) for the opportunity to discuss this and other related issues on elastic and diffractive scattering.

L. J. was supported by the Ukrainian Academy of Sciences' program "Structure and dynamics of statistical and quantum systems".

\end{document}